\documentclass[12pt]{article}
\usepackage{epsf,a4}

\newcommand{\Rsub}{\rm\scriptscriptstyle}

\begin{document}

\title{The Static Potential in QCD -- a full Two-Loop Calculation%
\footnote{Work supported in part by Graduiertenkolleg ``Elementarteilchenphysik
an Beschleunigern'', by the ``Landesgraduiertenf\"orderung'' at the
University of Karlsruhe, and by BMBF under contract 057KA92P.}}
\author{Markus Peter \\ 
 {\em \normalsize Institut f\"ur Theoretische Teilchenphysik,}\\[-0.2cm]
 {\em \normalsize  Universit\"at Karlsruhe}\\[-0.2cm] 
 {\em \normalsize  D--76128 Karlsruhe, Germany}}
\date{February 4, 1997 \\ revised May 16, 1997}
\maketitle
\thispagestyle{empty}
\vspace{-4.0truein}
\begin{flushright}
{\bf TTP 97-03\footnote{
    The complete paper, including figures, is
    also available via anonymous ftp at
    ttpux2.physik.uni-karlsruhe.de (129.13.102.139) as
    /ttp97-03/ttp97-03.ps, or via www at
    http://www-ttp.physik.uni-karlsruhe.de/cgi-bin/preprints.}
}\\
{\bf February 1997}\\
{\bf hep-ph/9702245}
\end{flushright}
\vspace{3.0truein}
\begin{abstract}
  A full analytic calculation of the two-loop diagrams contributing to
  the static potential in QCD is presented in detail. Using a
  renormalization group improvement, the ``three-loop'' potential in
  momentum space is thus derived and the third coefficient of the
  $\beta$-function for the $V$-scheme is given. The Fourier transformation
  to position space is then performed, and the result is briefly discussed.
\end{abstract}

\section{Introduction}
The famous Coulomb potential of electrodynamics is very important as
an essential ingredient in any non-relativistic problem involving
charged particles and consequently as being responsible for many
phenomena in everyday life. It is thus no surprise that its analogue
in chromodynamics has also been of great interest since 20 years.
Although the QCD potential certainly is not as ubiquitous in the
macroscopic world, it still represents a fundamental concept which,
besides the fact that potential models have been astonishingly
successful in the description of quarkonia, might give us some deeper
understanding of non-abelian theories and especially of confinement.
There is, of course, no known way to analytically derive the confining
part of the potential from first principles up to now, and
consequently this paper will also be restricted to an analysis of the
perturbative part. Nevertheless one may hope to obtain some hints
on the non-perturbative regime in this way. In addition, a calculation
of the perturbative potential is required for comparisons between
continuum QCD and lattice results.

The first investigations of the interaction energy between an
infinitely heavy quark-antiquark pair date back to 1977
\cite{Susskind,F77,ADM77} and were initiated by L. Susskind's Les Houches
lectures \cite{Susskind}. In these works some general properties of
the static potential were discussed and the one-loop and parts of the
higher-order diagrams were computed. In the years that followed, much
effort went into deriving (quark-mass suppressed) spin- and
velocity-dependent corrections in various frameworks, but it is
surprising that for a very long time there was no complete two-loop
calculation of the true static case available. Only recently has this
gap been closed \cite{MP96}, and the present paper is essentially an
extended version of \cite{MP96} presenting details about the two-loop
calculation, which might be of interest for other problems as well.

Before turning to the real problem, however, a brief description of the
structure of the paper should not be missing: as an introduction of the
notation and for illustrative purposes, the QED potential is discussed in
section \ref{sec:QED}, which is followed by a section describing the main
additional problems encountered in the non-abelian theory. Section
\ref{sec:Tech} is concerned with the techniques needed for the actual two-loop
calculation, and finally sections \ref{sec:Res} and \ref{sec:pos} contain the
result for the potential in momentum and position space, respectively.

\section{Abelian case: the QED potential \label{sec:QED}}

A definition of the static potential in QED which is both convenient
for analytical and lattice calculations and which in addition is
manifestly gauge invariant can be given as follows:\\
\parbox{10cm}{
  consider the vacuum expectation value of a rectangular Wilson loop
  of spatial extent $r$ and temporal extent $t$,
$$
    \langle 0|{\rm T}\;\exp\{ie\oint dx_\mu A^\mu\}|0\rangle.
$$
}\hspace{15mm}\parbox{35mm}{ \epsfxsize2.5cm \epsfbox{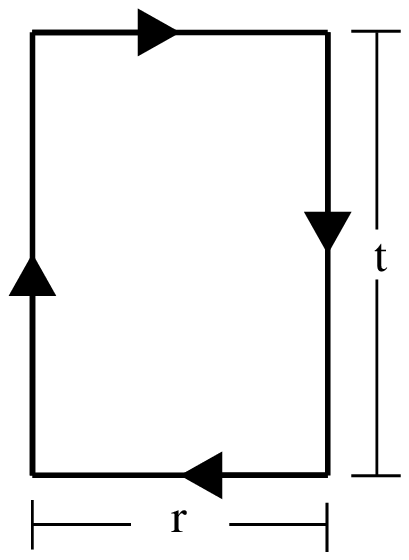} }\\
If we let $t$ approach infinity, the lines corresponding to $|t|=\infty$ will
only give a negligible contribution to the line integral because the length of
the integration path is much smaller than that of the other two
lines\footnote{It should be evident from the definition (\ref{def}) that a
  small contribution from these two lines would indeed
  not affect the potential.}.  In other words,
\begin{eqnarray*}\lefteqn{
  \langle 0|{\rm T}\;\exp\{ie\oint dx_\mu A^\mu\}|0\rangle}&& \\
 &\stackrel{t\to\infty}{\longrightarrow}&
  \langle 0|{\rm T}\;\exp\{ie\int_{-t/2}^{t/2}d\tau\Big(A_0(\tau,-{\bf r}/2)
    -A_0(\tau,{\bf r}/2)\Big)\}|0\rangle.
\end{eqnarray*}
The path integral representation of the remaining expectation value
equals the partition function of a system described by the
usual (purely photonic) QED Lagrangian with the addition of a source
term $J_\mu(x)A^\mu(x)$ with
$$
  J_\mu(x) = ev_\mu\Bigg[\delta^{(3)}\Big({\bf x}+\frac{\bf r}{2}\Big)
    - \delta^{(3)}\Big({\bf x}-\frac{\bf r}{2}\Big)\Bigg];\qquad
    v_\mu \equiv \delta_{\mu0}.
$$

As the partition function is dominated by the ground state energy in this
limit, i.e.\ the path integral approaches $\exp(-itE_0(r))$, and as the ground
state energy is exactly what we would term the potential, we are led to
define
\begin{equation}\label{def}
  V(r) = -\lim_{t\to\infty}\frac{1}{it}\ln\langle 0|{\rm T}\;\exp\{ie\oint
  dx_\mu A^\mu\}|0\rangle.
\end{equation}

For pure QED the vacuum expectation value is a gaussian path integral
and an exact calculation is therefore possible, with the result
\begin{equation}\label{QED}
  V(r) = e^2\int\!\frac{d^3q}{(2\pi)^3}\Bigg(\frac{1}{{\bf q}^2} -
    \frac{e^{i{\bf qr}}}{{\bf q}^2}\Bigg) = \Sigma+V_{\Rsub Coul.}
\end{equation}
where as the second term the expected Coulomb potential appears, but
there is also an infinite constant representing the self-energy of the
sources known from classical electrodynamics.

An exact solution will no longer be possible as soon as light fermions are
included or a non-abelian theory is considered. It is therefore useful to
perform a perturbative analysis as well. The Feynman rules for the sources can
be read off easily: each source-photon vertex corresponds to a factor
$iev^\mu$, an anti-source obtains an additional minus sign. When we expand the
time-ordered exponential, we may introduce $\Theta$-function to express the
different possible time-orderings of the fields $A^\mu$, and in turn can
re-interpret them as source propagators in position space,
\begin{equation}
  S_F(x-x^\prime) = -i\Theta(x_0-x_0^\prime)\delta^{(3)}({\bf x-x}^\prime),
\end{equation}
the reverse time ordering must be used for the anti-source.
Transforming the expression to momentum space one obtains
\begin{equation}
  S_F(p) = \frac{1}{vp+i\varepsilon}
\end{equation}
with $v\to-v$ for the anti-source.  One should note that in the path-integral
approach we could in fact omit the time-ordering prescription, as it is
already implicit in that formalism, but the final integrals we would encounter
would not be easier to solve. The approach we take, however, will prove
to be very useful in the non-abelian theory.

In addition, an immediate observation is that the Feynman rules are exactly the
same as those of Heavy Electron Effective Theory, the QED analogue of HQET,
where $v$ would represent the electron's velocity. This comes as no surprise,
as we are investigating the infinite mass limit of QED, and it is well known
that the potential can also be derived from the scattering operator. The static
QED potential thus should be derivable from the scattering matrix of HEET,
correspondingly the QCD potential from HQET, and this is in fact what is done
in practice. This approach can even be used to determine the spin-dependent
corrections to the potential \cite{CKO95}.

\begin{figure}\begin{center}
 \epsfxsize 8cm \mbox{\epsfbox{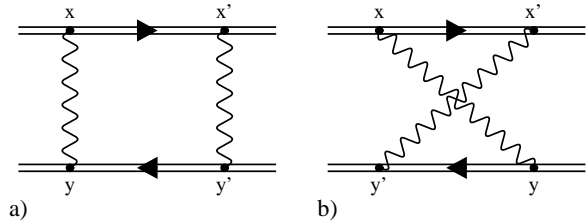}}
 \end{center}
 \caption{One-loop ladder diagrams. The double-lines represent the static
 sources.\label{fig:one}}
\end{figure}

At tree level, one obtains the scattering amplitude
$$ 
   -ie^2 \int\!\!dx_0dy_0\;v_\mu v_\nu D^{\mu\nu}(x-y)\Big|_{{\bf x}=0,{\bf y}
     ={\bf r}}= 
   -ie^2t\int\!\frac{d^3q}{(2\pi)^3}\frac{e^{i{\bf qr}}}{-{\bf q}^2}
$$
where $D^{\mu\nu}$ is the photon propagator. At first glance this seems to be
the complete result already and one might wonder what happens with the
higher order diagrams. But some care is required because the sum of all
diagrams does not correspond to the potential, but to its exponential. That
is exactly what the loop graphs are needed for. At one-loop order
for example, the ladder diagrams shown in Fig.\ \ref{fig:one} appear. When
working in position space it is easy to see that summing the two effectively
removes the anti-source propagator --- because it is only a
$\Theta$-function.  Hence, adding them once more with $x\leftrightarrow
x^\prime$ the source propagator is also removed and one finds
$$
 2!\times\mbox{Fig.}~\ref{fig:one}
  = \Big(-ie^2t \int\!\!dx_0dy_0\;D^{00}(x-y)\Big|_{{\bf x}=0,{\bf y}
    ={\bf r}}\Big)^2.
$$
(A similar analysis in momentum space is also feasible, but more difficult
\cite{MP96}.)  It is clear from the way the ladder diagrams are constructed
that this behaviour persists to all orders: the ladder diagrams are derived
from the uncrossed ladder by permuting all vertices, i.e.\ generating all
time-orderings, on one of the two fermion lines. Hence after summing all of
them we end up with the uncrossed ladder again, where the $\Theta$-functions on
one of the fermion lines are removed. The source propagators on the other line
are removed by adding the same sums, with all possible permutations of the
names of the vertices, just as was done in the one-loop example. Since there
are $n!$ permutations at $n$-loops, the exponential thus forms.

The other types of graphs --- source self-energy and vertex correction
diagrams --- can be shown, along the same line, to produce the
products of the form $\Sigma^n\cdot V^m_{\Rsub Coul.}(r)$ predicted by
the exact formula (\ref{QED}).

If light fermions are to be included, an explicit exact expression for the
potential can no longer be derived and one has to rely on the perturbative
treatment. The effect of the fermions is to introduce a running coupling in
two different ways. The obvious point is that fermion loops cause vacuum
polarization effects, which can be accounted for by defining an effective
charge in momentum space via
$$
  \alpha_{\Rsub eff}({\bf q}^2) = \frac{\alpha}{1+\Pi({\bf q}^2)}
$$
where $\Pi({\bf q}^2)$ represents the vacuum polarization function and
$\alpha$ the fine structure constant. This definition is gauge
invariant and includes a Dyson summation. One might thus guess that
\begin{equation}\label{QEDeff}
  V({\bf q}^2) = -\frac{4\pi\alpha_{\Rsub eff}({\bf q}^2)}{{\bf q}^2},
\end{equation}
but the formula need only be correct at the one- and two-loop level,
because starting at three loops light-by-light scattering diagrams enter.
The correct formula should read
\begin{equation}\label{QEDV}
  V({\bf q}^2) = -\frac{4\pi\alpha_{\Rsub V}({\bf q}^2)}{{\bf q}^2},
\end{equation}
with $\alpha_{\Rsub V}\neq\alpha_{\Rsub eff}$, which, however, merely serves
as a definition of $\alpha_{\Rsub V}$. Nevertheless, this definition is
quite convenient, especially when turning to the non-abelian case where
a gauge-invariant definition of an effective charge is nontrivial.

\section{\label{sec:QCD}Complications in the non-abelian case}

One novelty that arises in QCD is well known: the non-linear nature of
non-abelian theories introduces additional diagrams which cause a running
coupling even in the absence of light fermions, and the fact that gluons carry
colour also requires a redefinition of the Wilson loop and consequently
of the potential as follows:
\begin{equation}
  V(r) = -\lim_{t\to\infty}\frac{1}{it}\ln
    \langle0|\mbox{Tr~P}\exp\Big(ig \oint\!dx_\mu\;A^\mu_aT^a \Big)|0\rangle.
\end{equation}
The generators $T^a$ have to be inserted in order to absorb the gluons' colour
index, and the fact that the generators do not commute requires the
introduction of a path-ordering prescription. An important point is that this
prescription is not automatically taken into account by using the path-integral
formalism, in contrast to the time-ordering.  Consequently, all we can do even
in the path-integral approach is use the definition of the path-ordered
exponential, i.e.\ expand it, and the usefulness of the way we analyzed the
abelian case should become clearer. The net effect of the modifications in the
non-abelian case is a complication of the way the exponentiation of the
potential arises, and that the sources are automatically in a colour singlet
state.

In principle we are free to choose any representation for the generators, but
since we intend to describe quarks we are going to use the fundamental one. On
tree level, the only change with respect to QED then is the colour factor $C_F
= T_F(N^2-1)/N$ that multiplies the coupling in the potential, where $N$ is
the number of colours and $T_F$ the normalization of the generators. It is
thus convenient to define, in analogy to (\ref{QEDV}),
\begin{equation}\label{QCDV}
  V({\bf q}^2) = -C_F\frac{4\pi\alpha_{\Rsub V}({\bf q}^2)}{{\bf q}^2},
\end{equation}
as this has the following advantage: $C_F$ is the value of the Casimir
operator $T^aT^a=C_R${\bf 1} in the fundamental representation
$T^a_{ij}=\lambda^a_{ij}/2$, where the $\lambda^a$ are the Gell-Mann matrices.
If we chose the adjoint representation $T^a_{ij}=-if^{aij}$, which would be
equally acceptable, the only thing that would change in (\ref{QCDV}) is the
overall factor $C_F$ which would be replaced by $C_A$, the coupling
$\alpha_{\Rsub V}$ would remain the same.  The statement is trivial at tree
level, but it is in fact true to all orders as will be explained after
the discussion on the exponentiation. By choosing the adjoint representation
we obtain the static potential for gluinos in a colour singlet state.

In higher orders the presence of the generators causes the individual
diagrams to obtain different colour factors and consequently the
exponentiation has to be more involved than in the abelian case.
Although this point has already been analyzed in \cite{F77}, it seems
appropriate to recall the discussion here. It should, however, suffice to
consider the ladder diagrams to explain the basic idea.

The QCD one-loop ladder diagrams corresponding to Fig.\ \ref{fig:one}
obtain the colour factors
\begin{equation}
  \mbox{Fig.\ \ref{fig:one}a}\propto C_F^2 \qquad,\qquad
  \mbox{Fig.\ \ref{fig:one}b}\propto C_F^2-\frac{C_A}{2}C_F
\end{equation}
and we immediately can identify the abelian-like terms $\propto C_F^2$
that are needed to build the exponential of the Coulomb potential. But
it is also obvious that the remainder of Fig.\ \ref{fig:one}b,
together with a contribution from the vertex corrections involving the
same colour factor, gives an additional contribution to the potential,
which then has to be exponentiated by the higher order diagrams.

\begin{figure}[ht]\begin{center}
 \epsfxsize 12cm \mbox{\epsfbox{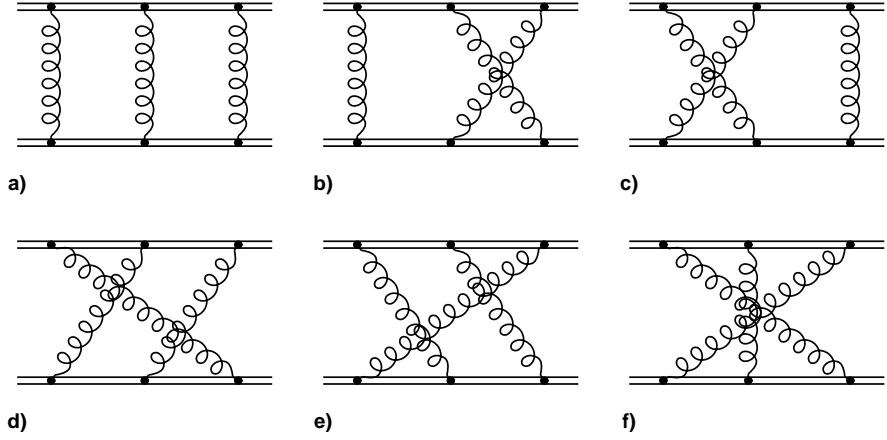}}
 \end{center}
 \caption{Two-loop ladder diagrams in QCD.\label{fig:two}}
\end{figure}

Fig.\ \ref{fig:two} shows the two-loop ladder diagrams, which involve the
colour factors
\begin{eqnarray}
  \mbox{Fig.}~\ref{fig:two}a~~ & \propto & C_F^3 \\
  \mbox{Fig.}~\ref{fig:two}b,c & \propto & C_F^3 - C_F^2\frac{C_A}{2} 
    \label{tlbc}\\
  \mbox{Fig.}~\ref{fig:two}d,e & \propto & C_F^3 - 2C_F^2\frac{C_A}{2}
    +\frac{C_A^2}{4}C_F\label{tlde}\\
  \mbox{Fig.}~\ref{fig:two}f~~ & \propto & C_F^3 - 3C_F^2\frac{C_A}{2}+
    \frac{C_A^2}{2}C_F.\label{tlf}
\end{eqnarray}
As expected, each diagram contains the ``abelian'' term $C_F^3$, thus forming
an iteration of the tree-level potential, but the terms $\propto C_F^2C_A$ are
also iterations, exactly those that arise from exponentiating the additional
one-loop term. This can be seen as follows: when working in coordinate space
and neglecting the colour factors, we can look for combinations of the diagrams
in which one of the anti-source propagators is removed, with the result
\begin{eqnarray*}
  (b)+(e)+(f) &=&\raisebox{-0.35in}{\epsfxsize1.25in 
       \epsfbox{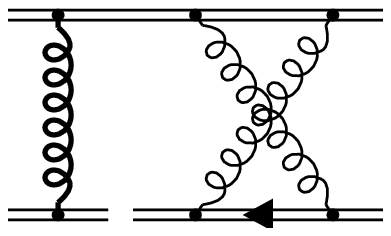}}  \\
  (c)+(d)+(f) &=&\raisebox{-0.35in}{\epsfxsize1.25in
       \epsfbox{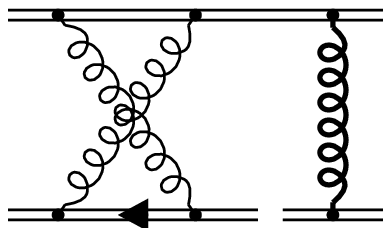}} \\
  (d)+(e)+(f) &=&\raisebox{-0.35in}{\epsfxsize1.25in 
       \epsfbox{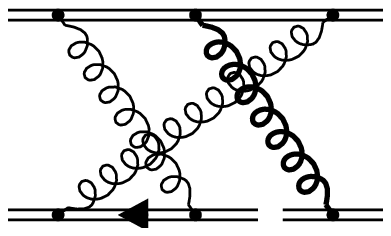}}.
\end{eqnarray*}
Summing the three lines we find that the coefficients of $-C_F^2C_A/2$ in
Eqs.\ (\ref{tlbc}--\ref{tlf}) equal the corresponding diagrams' coefficients in
the sum and that the $\Theta$-functions fixing the position of the marked
gluon line are removed. Hence the sum is a product of one-gluon exchange and
the crossed ladder diagram with exactly the right colour factor. A similar
analysis for the remaining two-loop diagrams leads to the conclusion that the
exponentiation works in the non-abelian case as well, and that all terms
$\propto C_F^3$ or $C_F^2C_A$ indeed constitute iterations.

In consequence, the number of diagrams that really have to be computed
to determine the two-loop contribution to the QCD potential is
somewhat reduced. Nevertheless, when compared to QED the number of diagrams is
still quite large, as in the abelian case only very few remain
(essentially the vacuum polarization).

Having discussed the exponentiation we are able to proof the statement
that $\alpha_{\Rsub V}$ is independent of the representation the sources are
in. It is obvious that for any representation $R$, the colour factor for a
given diagram is the same as the one for $R=F$ if we replace $C_F$ by $C_R$,
because it is completely determined by the algebra of the generators: only
$C_R$ and $C_A$ can appear since the generators themselves and the structure
constants arising from commutators are the only ``matrices'' around, and their
combination is fixed. We have also seen that at each loop order the only net
contribution to the potential arises from the colour factors linear in $C_F$,
all other terms are iterations. The definition (\ref{QCDV}) then implies that
$\alpha_{\Rsub V}$ does not involve $C_F$ at all%
\footnote{We only consider factors $C_F$ coming from the sources here. Of
  course $\alpha_{\Rsub V}$ does involve $C_F$, but these terms arise from
  fermion loops and are unrelated to the representation of the external
  sources. Consequently they should not be replaced.}
and thus the replacement $C_F\to C_R$ does not affect the coupling.

\section{\label{sec:Tech}Details of the two-loop calculation}

It is worth presenting at least some details of the techniques used to
compute the two-loop diagrams, first of all because they are helpful when
trying to reproduce the results, but also because they might prove useful
for other calculations as well.

The remaining diagrams are best analyzed in momentum space, using the
kinematics that follows from the Wilson loop definition: the ``on-shell
condition'' for the source reads $vp=0$ where $p$ denotes the four-momentum
carried by the source. Thus the sources may have any three-momenta, the
actual values of which are, however, unimportant, as the only quantity that
appears is the momentum transfer $q^\mu=(0,{\bf q})$.

By choosing a convenient gauge the number of graphs can be further reduced:
if we employ Feynman gauge, all diagrams with a three- or four gluon vertex
with all gluons directly coupled to source lines vanish. This welcome feature
is caused by the replacement of the Dirac matrices $\gamma^\mu$ in the
source-gluon vertex by the simple factor $v^\mu=\delta^{\mu0}$, which means
that all vertices are interchangeable as far as their Lorentz structure and
momentum dependence is concerned (the momentum dependence is mentioned
because it is the property which is destroyed by other gauges). The
symmetry properties of both the three- and four gluon vertices then imply
that the diagrams vanish.

\begin{figure}[ht]\begin{center}
 \epsfxsize 12cm \mbox{\epsfbox{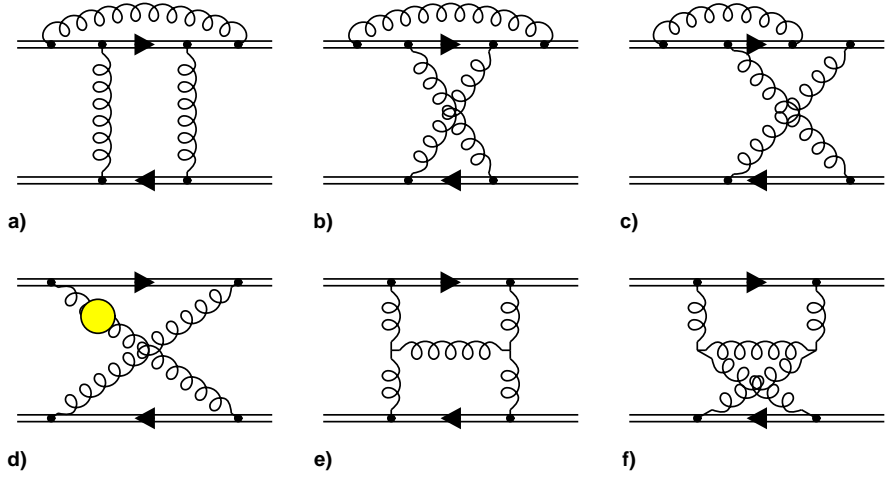}}
 \end{center}
 \caption{Two-gluon exchange diagrams in QCD that remain in Feynman gauge.
  \label{fig:rest}}
\end{figure}

Consequently, apart from the ladders Fig.\ \ref{fig:two}d--f, the two-loop
gluon self-energy and double insertions of one-loop corrections, only a few of
the two-loop vertex correction graphs and the six two-gluon exchange diagrams
of Fig.\ \ref{fig:rest} have to be calculated. In fact the number of diagrams
is even slightly lower because Fig.\ \ref{fig:rest}f has a vanishing colour
factor for sources in the colour singlet state, and of the ladder diagrams
only Fig.\ \ref{fig:two}f is required, because the relation
\begin{equation}
  \mbox{Fig.~\ref{fig:two}d+\ref{fig:two}e} = - \mbox{Fig.~\ref{fig:two}f}
\end{equation}
holds if the corresponding colour factors are neglected. The equality
can easily be proven in momentum space by applying the trivial identity
$$
   \frac{1}{lv-kv+i\varepsilon}\frac{1}{kv+i\varepsilon} -
   \frac{1}{lv-kv+i\varepsilon}\frac{1}{lv+i\varepsilon} =
     \frac{1}{kv+i\varepsilon}\frac{1}{lv+i\varepsilon}
$$
to one of the two source lines.

Using dimensional regularization with $D=4-2\epsilon$ for both the infrared
and ultraviolet divergencies, the integration by parts technique \cite{IbP}
can be used to reduce most of the integrals that arise to products or
convolutions of four types of one-loop integrals, that in turn can be
computed by standard methods. With the notation
$$ \widetilde{dp} \equiv \frac{(2\pi\mu)^{2\epsilon}}{i\pi^2}d^Dp\qquad,\qquad
   {\cal P}_n(p) \equiv p^{\mu_1}p^{\mu_2}\cdots p^{\mu_n},
$$
\begin{equation}
 G(n,m;k,j) = (4\pi\mu^2)^\epsilon\frac{\Gamma(n+m-j-{D\over2})}
 {\Gamma(n)\Gamma(m)}B\Big(\frac{D}{2}-n+k-j,\frac{D}{2}-m+j\Big),
\end{equation}
\begin{equation}
 G_H(n,m;k,j) = (4\pi\mu^2)^\epsilon\frac{\Gamma(\frac{D}{2}-n+k-j)
   \Gamma(2n+m-k-D)}{\Gamma(n)\Gamma(m)}
\end{equation}
the integrals are, omitting the $i\varepsilon$ in the propagators for
brevity:
\begin{eqnarray}
  \lefteqn{
   \int\widetilde{dp}\Bigg(\frac{1}{-p^2}\Bigg)^n\Bigg(\frac{1}{-(p-q)^2}
   \Bigg)^m {\cal P}_k(p)} && \nonumber \\ &=& 
     (-q^2)^{D/2-n-m}\sum_{j=0}^{[k/2]}G(n,m;k,j)
     \Big(\frac{q^2}{4}\Big)^j\frac{1}{j!}\Bigg(\frac{\partial}{\partial q_\mu}
     \frac{\partial}{\partial q^\mu}\Bigg)^j{\cal P}_k(q),\label{two:std}\\
  \lefteqn{
    \int\widetilde{dp}\Bigg(\frac{1}{-p^2}\Bigg)^n\Bigg(\frac{w}{pv+w}
    \Bigg)^m {\cal P}_k(p)} && \nonumber \\ &=&
    (-1)^k(-2w)^{D+k-2n}\sum_{j=0}^{[k/2]}G_H(n,m;k,j)\Big(\frac{-1}{4}\Big)^j
    \Bigg(\frac{\partial}{\partial v_\mu}\frac{\partial}{\partial v^\mu}
    \Bigg)^j{\cal P}_k(v),\label{two:hqet}\\
  \lefteqn{
    \int\widetilde{dp}\Bigg(\frac{1}{-p^2}\Bigg)^n\Bigg(\frac{1}{-(p-q)^2}
   \Bigg)^m\Bigg(\frac{1}{-pv}\Bigg)^a\Bigg|_{vq=0}} && \nonumber \\ &=&
   (-q^2)^{\frac{D-a}{2}-n-m}\frac{\sqrt\pi}{\Gamma(\frac{a+1}{2})}
   G(n,m;-a,-a/2),\label{three:one}\\
  \lefteqn{
    \int\widetilde{dp}\Bigg(\frac{1}{-p^2}\Bigg)^n\Bigg(\frac{w}{pv+w}
    \Bigg)^a\Bigg(\frac{w}{pv}\Bigg)^b} && \nonumber \\ &=&
     (-2w)^{D-2n}\frac{\Gamma(D-2n-b)}{\Gamma(D-2n)}G_H(n,a;-b,-b).
     \label{three:two}
\end{eqnarray}
The formula for the standard massless two-point function Eq.\ (\ref{two:std})
is a well known result which can be found for example in \cite{IbP}. Its HQET
analogue (\ref{two:hqet}) and the HQET three-point functions (\ref{three:one})
and (\ref{three:two}) can be calculated in a similar way using the
parametrization
$$ \frac{1}{a^n b^m} = \frac{\Gamma(n+m)}{\Gamma(n)\Gamma(m)}\int_0^\infty\!
   d\alpha \frac{\alpha^{m-1}}{(a+\alpha b)^{n+m}}
$$
to combine gluon and source propagators, instead of the usual Feynman
parameters. Eq.\ (\ref{two:hqet}) without the factor ${\cal P}_k(p)$ has
already been given in \cite{BG91}, and Eq.\ (\ref{three:two}) can be derived
from Eq.\ (8) of \cite{BBG93}.

The reduction of a given graph to the two functions $G$ and $G_H$ can be
automated with the help of a computer program such as FORM \cite{Form}, but
an additional final reduction to a few basic $G_{(H)}$-functions seems
quite difficult because of the appearance of $\epsilon$-dependent values for
the last two arguments of $G$ and $G_H$ and additional ratios of
$\Gamma$-functions.  As an example consider the integral
\begin{eqnarray*}
  \lefteqn{
    \int\!\widetilde{dp}\widetilde{dp^\prime}\frac{1}{p^2(p+q)^2(p^\prime)^2
    (p^\prime v)^2(p^\prime v+pv)(pv)}} && \\
  &\stackrel{(\ref{three:two})}{=}&
    -\frac{2^{D-2}\Gamma(D-4)}{\Gamma(D-2)}G_H(1,1;-2,-2)
    \int\!\widetilde{dp}\frac{1}{(-p^2)(-(p+q)^2)(-pv)^{6-D}} \\
  &\stackrel{(\ref{three:one})}{=}& 
    \frac{-8\Gamma(-2\epsilon)\Gamma(1+\epsilon)}{\Gamma(2-2\epsilon)
    \Gamma(2+2\epsilon)}G_H(1,1;-2,-2)G(1,1;-2-2\epsilon,-1-\epsilon)
    (-q^2)^{-1-2\epsilon}
\end{eqnarray*}
which corresponds to Fig.\ \ref{fig:rest}b. It thus seems easier to directly
rewrite the resulting expressions in terms of $\Gamma$-functions and
immediately expand them in $\epsilon$.

However, as already mentioned, not all of the graphs can be reduced in an
easy way to the above one-loop integrals, a few diagrams remain that involve
the following type of true two-loop integrals:
\begin{equation}
  I(a,b,c;n,m) = \int\!\widetilde{dp}\widetilde{dr}\Bigg(\frac{-1}{p^2}\Bigg)^a
  \Bigg(\frac{-1}{r^2}\Bigg)^b\Bigg(\frac{-1}{(p-r-q)^2}\Bigg)^c
  \frac{1}{(pv)^n}\frac{1}{(rv)^m}
\end{equation}
with $vq=0$ and $n,m>0$. From
$$
  \int\!\widetilde{dp}\widetilde{dr}\;v_\mu\frac{\partial}{\partial p_\mu}
    f(p,r;q) = 0
$$
one can derive the recursion relation
\begin{equation}
  n{\rm N}^+I = 2(a{\rm A}^+ + c{\rm C}^+){\rm N}^-I - 2c{\rm C}^+{\rm M}^- I.
\end{equation}
Here the operator ${\rm N}^+$ means that the argument $n$ of $I$ should be
increased by one, and correspondingly for the other operators. The relation
can be used to reduce integrals with $n>1$ or $m>1$ to those with $n=m=1$. For
example, with $n=1,m=2$ we find for the integral corresponding to Fig.\ 
\ref{fig:two}f:
\begin{equation}
  I(1,1,1;2,2) = 2I(2,1,1;0,2)+2I(1,1,2;0,2)-2I(1,1,2;1,1).
\end{equation}
The first two terms can be computed with the help of
(\ref{two:std})--(\ref{three:two}), but the last one cannot be simplified any
further by the recursion relation (due to the absence of the propagators
$1/(p-q)^2$ and $1/(r-q)^2$ there are no other simple relations).

It turns out that the following three ``irreducible'' integrals,
the calculation of which is sketched in the appendix, are needed:
\begin{eqnarray}
  I(1,1,1;1,1) & = & \frac{2\pi}{3}(-q^2)^{-2\epsilon}G(1,1;-1,-\frac{1}{2})
     G(1,\frac{1}{2}+\epsilon;-1,-\frac{1}{2})\\
  I(1,1,2;1,1) & = & -\frac{2}{q^2}\Bigg(\frac{4\pi\mu^2}{-q^2}\Bigg)^{
    2\epsilon}\Bigg[\frac{1}{\epsilon^2}-\frac{2}{\epsilon}+4-\frac{5}{6}\pi^2
     \nonumber \\ &&
    -\epsilon\Big(8-\frac{5}{3}\pi^2+\frac{32}{3}\zeta_3\Big)
    +{\cal O}(\epsilon^2)\Bigg]\\
  I(2,1,2;1,1) & = & \frac{1}{(q^2)^2}\Bigg(\frac{4\pi\mu^2}{-q^2}\Bigg)^{
    2\epsilon}\Bigg[\frac{2}{\epsilon^2}+\frac{2}{\epsilon}-4-\frac{5}{3}\pi^2
     \nonumber \\ &&
    -\frac{\epsilon}{3}\Big(24-17\pi^2-64\zeta_3\Big)
    +{\cal O}(\epsilon^2)\Bigg].
\end{eqnarray}

With these equations all the basic formul\ae\ for the determination of
the two-loop diagrams are given. The expressions for the individual graphs
will, however, not be listed here, we will directly turn to the complete
result instead.

\section{\label{sec:Res}The QCD potential in momentum space}

A convenient way of writing the QCD potential in momentum space is\footnote{
  The second and third equations might be incomplete for $n>2$, because $a_n$
  might depend on $\ln\alpha$, see the discussion in \cite{ADM77}. The present
  paper, however, is restricted to the three-loop potential, i.e.\ $n\le2$.}
\begin{eqnarray}
 V({\bf q}^2) & = & -C_F\frac{4\pi\alpha_{\Rsub V}({\bf q}^2)}{{\bf q}^2}
   \label{Vq}\\
 \alpha_{\Rsub V}({\bf q}^2) & = & \alpha_{\overline{\Rsub MS}}(\mu^2)
   \sum_{n=0}^\infty \tilde{a}_n(\mu^2/{\bf q^2})
     \Big(\frac{\alpha_{\overline{\Rsub MS}}(\mu^2)}{4\pi}\Big)^n
   \label{orig} \\ &=& 
   \alpha_{\overline{\Rsub MS}}({\bf q}^2)
   \sum_{n=0}^\infty a_n\Big(\frac{\alpha_{\overline{\Rsub MS}}({\bf q}^2)}
   {4\pi}\Big)^n \label{vms}
\end{eqnarray}
with $a_0=\tilde a_0=1$ and where
\begin{equation}
  a_1 = \frac{31}{9}C_A - \frac{20}{9}T_Fn_f\quad,\quad
  \tilde a_1 = a_1 +\beta_0\ln\frac{\mu^2}{\bf q^2}
\end{equation}
is the long-known one-loop result and
\begin{eqnarray}
  a_2 &=& \Big(\frac{4343}{162}+6\pi^2-\frac{\pi^4}{4}+\frac{22}{3}\zeta_3
    \Big)C_A^2 -\Big(\frac{1798}{81}+\frac{56}{3}\zeta_3\Big)C_AT_Fn_f
    \nonumber \\ &&
    -\Big(\frac{55}{3}-16\zeta_3\Big)C_FT_Fn_f + \Big(\frac{20}{9}T_Fn_f\Big)^2
  \\
  \tilde a_2 &=& a_2 + \beta_0^2\ln^2\frac{\mu^2}{\bf q^2}
   +(\beta_1+2\beta_0 a_1)\ln\frac{\mu^2}{\bf q^2}
\end{eqnarray}
is the new information added by the present analysis. Note that the last term
of $a_2$ could have been predicted from the one-loop result, it is exactly the
contribution that would be Dyson-summed in QED by introducing an effective
coupling. The third term originates from the two-loop vacuum polarization and
thus would also be included in the effective coupling in the case of QED.

Knowledge of $a_2$ now allows us to consistently use the three-loop
expression for the running coupling in the $\overline{\rm MS}$-scheme. An
equation of the form (\ref{vms}) can then be used to derive the
$\beta$-function of one of the two couplings from the knowledge of the
$\beta$-function of the second scheme.  If we define $\beta$ and the
coefficients $\beta_n$ in each scheme via
\begin{equation}
  \frac{1}{\alpha(\mu^2)}\frac{d\alpha(\mu^2)}{d\ln\mu^2}
  = - \beta(\alpha)
  = - \sum_{n=0}^\infty \beta_n\Big(\frac{\alpha(\mu^2)}{4\pi}\Big)^{n+1},
\end{equation}
we immediately find
\begin{equation}
  \beta^{\rm V}(\alpha_{\Rsub V}) = \beta^{\overline{\rm MS}}(
  \alpha_{\overline{\Rsub MS}})
  \frac{\alpha_{\overline{\Rsub MS}}}{\alpha_{\Rsub V}}
  \frac{d\alpha_{\Rsub V}}{d\alpha_{\overline{\Rsub MS}}}\qquad
  \mbox{with}~\alpha_{\overline{\Rsub MS}} 
    = \alpha_{\overline{\Rsub MS}}(\alpha_{\Rsub V})
\end{equation}
or $\beta_{0,1}^{\rm V}=\beta_{0,1}^{\overline{\rm MS}}$ and
\begin{eqnarray}
  \beta_2^{\rm V} &=&
    \beta_2^{\overline{\rm MS}}-a_1\beta_1^{\overline{\rm MS}}
     + (a_2-a_1^2)\beta_0^{\overline{\rm MS}} \\
  & = & \Big(\frac{618+242\zeta_3}{9}+\frac{11(24\pi^2-\pi^4)
    }{12}\Big)C_A^3 \nonumber \\ &&
    -\Big(\frac{445+704\zeta_3}{9}+\frac{24\pi^2-\pi^4}{3}\Big)C_A^2T_F
    n_f \nonumber \\ &&
    + \frac{2+224\zeta_3}{9}C_A(T_Fn_f)^2 -\frac{686-528\zeta_3}{9}C_AC_FT_Fn_f
    \nonumber \\ &&
    +2C_F^2T_Fn_f+\frac{184-192\zeta_3}{9}C_F(T_Fn_f)^2.
\end{eqnarray}
Inserting the numbers one finds that the numerical value of $\beta_2^V$ is
considerably larger than that of $\beta_2^{\rm \overline{MS}}$, which, for
comparison, is given by the expression \cite{TVZ80}
\begin{eqnarray}
 \beta_2^{\rm \overline{MS}} & = & \frac{2857}{54}C_A^3+2C_F^2T_Fn_f
    -\frac{205}{9}C_AC_FT_Fn_f-\frac{1415}{27}C_A^2T_Fn_f \nonumber \\
   && +\frac{44}{9}C_F(T_Fn_f)^2+\frac{158}{27}C_A(T_Fn_f)^2.
\end{eqnarray}

Hence the coupling in the potential runs faster than the
$\overline{\rm MS}$-coupling.

\begin{figure}[ht]\begin{center}
 \epsfxsize 12cm \mbox{\epsfbox{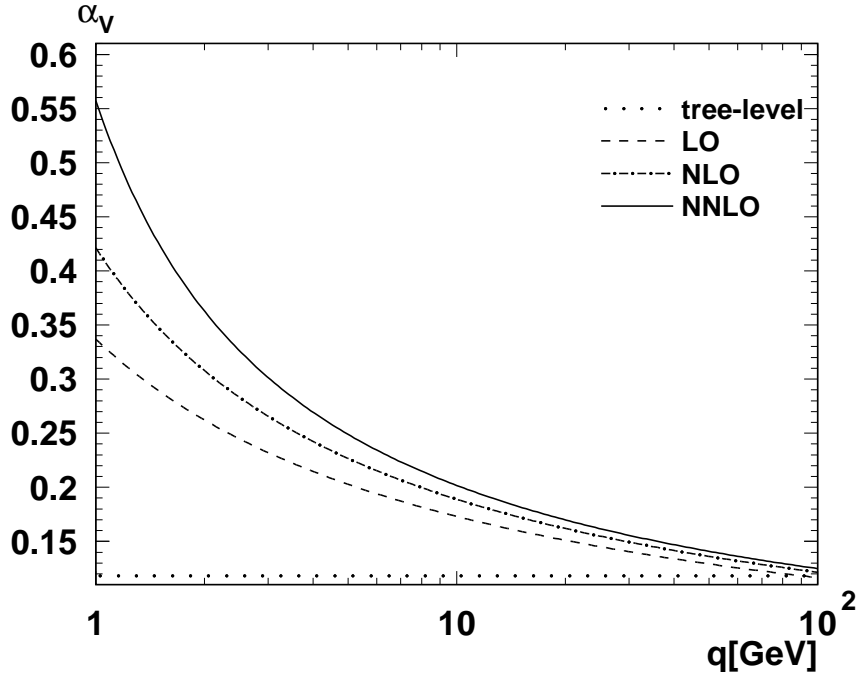}}
 \end{center}
 \caption{Comparison of the results for $\alpha_{\Rsub V}({\bf q}^2)$ at
  different loop-orders, where $\alpha_{\Rsub V}({\bf q}^2)$ is determined
  from $\alpha_{\overline{\Rsub MS}}({\bf q}^2)$. Input parameters are
  $\alpha_{\overline{\Rsub MS}}(M_Z^2)=0.118$, $n_f=5$.\label{fig:ava}}
\end{figure}

Figure \ref{fig:ava} compares the result for $\alpha_{\Rsub V}$ at the various
loop-orders in graphical form, neglecting quark thresholds for simplicity: the
dotted line represents the tree-level prediction, i.e.\ a pure Coulomb
potential without a running coupling; the dashed line shows the one-loop
result in the sense that only the (one-loop) running of
$\alpha_{\overline{\Rsub MS}}$ is taken into account, but the two couplings
still coincide (which means that it is still tree level as far as the diagrams
contributing to the potential are concerned and should thus better be termed
leading order); the dashed-dotted line shows the next-to-leading order and the
solid line the next-to-next-to-leading order results. It is evident that the
two-loop contribution is nearly as important as the one-loop effect: the
additional shift in $\alpha_{\Rsub V}$ caused by $a_2$ is roughly two third
the size of the shift introduced by $a_1$, and both corrections increase the
coupling. The impact of this shift on the would-be toponium system, as an
example, should be measurable: in the interval $|{\bf q|}=10\ldots30$GeV
including the NNLO-corrections amounts to a net increase of $\alpha_{\Rsub V}$
by about 5 to 7\%. The charmonium and bottomonium systems on the other hand
are mainly sensitive to momenta below 2GeV and thus lie predominantly outside
of the perturbative domain.

At this point we should mention that for simplicity we have chosen a fixed
number of five light flavours to be valid over the whole range of momenta. For
a more realistic study, the bottom and charm quark thresholds would, of
course, have to be taken into account by matching effective theories with
decreasing $n_f$ in order to restore decoupling. Such a procedure would,
however, leave our results qualitatively unchanged, it would mainly cause an
even faster running of both couplings at low momenta.

The plot just discussed was generated by employing (\ref{vms}) as it stands,
i.e.\ by evolving the $\overline{\rm MS}$-coupling, taken to equal $0.118$ at
the $Z$-mass, to the required scale (assuming $n_f=5$) and then calculating
$\alpha_{\Rsub V}$ at that scale. In principle the evolution can be
performed in different ways: one may express the coupling in terms of the
QCD scale parameter $\Lambda_{\Rsub QCD}$, or one may employ the QED-like
formula
\begin{eqnarray}\lefteqn{
  \frac{\alpha(\mu^2)}{\alpha(q^2)} = 1 + 
    \frac{\alpha(\mu^2)}{4\pi} \beta_0\ln\frac{q^2}{\mu^2} } && \nonumber\\ && 
   +\Big(\frac{\alpha(\mu^2)}{4\pi}\Big)^2 \beta_1\ln\frac{q^2}{\mu^2}
   +\Big(\frac{\alpha(\mu^2)}{4\pi}\Big)^3 \Big( \beta_2\ln\frac{q^2}{\mu^2}
     -\frac{1}{2}\beta_0\beta_1\ln^2\frac{q^2}{\mu^2} \Big).
\end{eqnarray}
Although the two approaches are formally equivalent, in practice only the
second one guarantees that the running coupling really coincides with its
input value $\alpha(\mu^2)$ for $q^2=\mu^2$. Therefore the second approach
has been adopted throughout this paper.

\begin{figure}[ht]\begin{center}
 \epsfxsize 12cm \mbox{\epsfbox{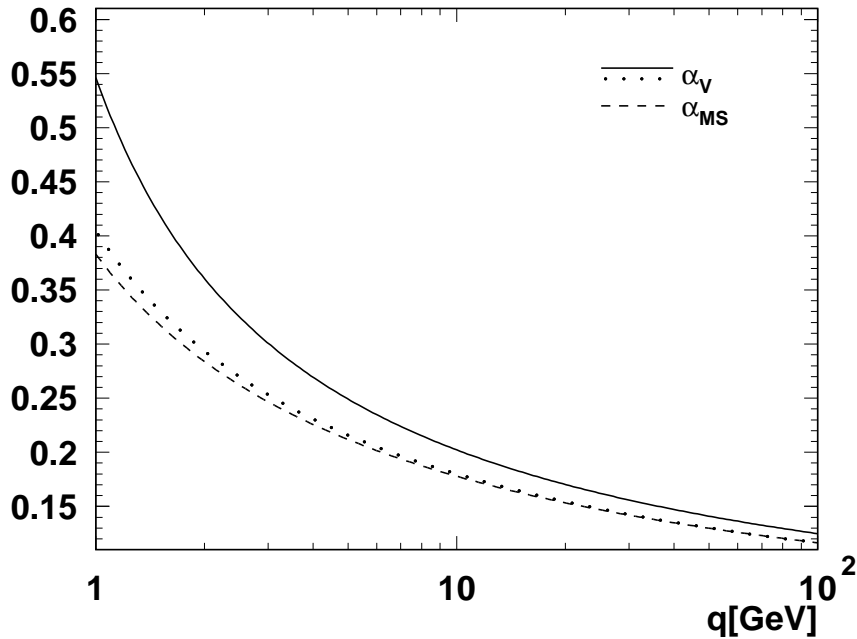}}
 \end{center}
 \caption{Comparison of the running of $\alpha_{\Rsub V}({\bf q}^2)$ (solid
  line) and $\alpha_{\overline{\Rsub MS}}({\bf q}^2)$ (dashed line)
  at three loops. The dotted curve displays $\alpha_{\Rsub V}({\bf q}^2)$
  with the initial value $\alpha_{\Rsub V}(M_Z^2)=\alpha_{\overline{\Rsub MS}}
  (M_Z^2)$\label{fig:rge}}
\end{figure}

One may argue that to first evolve the $\overline{\rm MS}$-coupling and then
determine the potential is not the best idea because, as is evident from
Fig.\ \ref{fig:ava}, the expansion parameter becomes large at low scales and
thus the next terms of the perturbation series might also become important.
An alternative would be to determine $\alpha_{\Rsub V}$ from (\ref{vms}) at
the $Z$-mass and then evolve it to the scale ${\bf q}^2$. It turns out, not
surprisingly, that the result is essentially the same, as can be seen from
Figure \ref{fig:rge} where the running of the two couplings is compared: the
solid line shows the three-loop running of $\alpha_{\Rsub V}$ and the dashed
one $\alpha_{\overline{\Rsub MS}}$, again for $\alpha_{\overline{\Rsub
    MS}}(M_Z^2)=0.118$ and five flavours; the dotted line shows the
corresponding result for $\alpha_{\Rsub V}$ for the hypothetical case
$\alpha_{\Rsub V}(M_Z^2)=0.118$ as well, and thus really displays the different
$\beta-$functions. It exemplifies that the large numerical mismatch
between the two couplings at low scales mainly arises from an amplification
of the small mismatch at large scales. Note that in a sense the two
couplings are the same at the two-loop level as the $\beta$-functions then
coincide and a difference merely arises because the evolutions start at
different values.

\section{\label{sec:pos}Position space}

With the momentum-space representation of $V_{\Rsub QCD}$ at our disposal, we
are able to compute the real analogue of the Coulomb potential, i.e.\ the QCD
potential in position space. In order to make the expressions simpler, it is
convenient to introduce the notation
\begin{equation} \label{Fu}
  {\cal F}(r,\mu,u) = \mu^{2u}
    \int\!\!\frac{d^3q}{(2\pi)^3}\frac{e^{i{\bf qr}}}{({\bf q}^2)^{1+u}}
\end{equation}
for the Fourier transform of a general power of $1/{\bf q}^2$. Using
a Schwinger parameter,
$$
   \frac{1}{({\bf q}^2)^{1+u}} = \frac{1}{\Gamma(1+u)}\int_0^\infty\!
   dx\;x^u\;e^{-x{\bf q}^2},
$$
and a few relations between $\Gamma$-functions such as
$$ 
   \Gamma(1+u) = \sqrt\pi \frac{\Gamma(1+2u)}{2^{2u}\Gamma(\frac{1}{2}+u)},
   \qquad
   \ln\Gamma(1+u) = \gamma_E u +\sum_{n=2}^\infty\frac{\zeta(n)}{n}u^n\quad
   \mbox{for}~|u|<1,
$$
several equivalent representations of ${\cal F}$ can be derived, two of which
will be useful in the following:
\begin{eqnarray}
  {\cal F}(r,\mu,u) & = & 
     \frac{(\mu r)^{2u}}{4\pi^2 r}\frac{\Gamma(\frac{1}{2}+u)\Gamma(\frac{1}{2}
       -u)}{\Gamma(1+2u)} \label{Fua} \\
  & = &
     \frac{(\mu r e^{\gamma_E})^{2u}}{4\pi r}\exp\Bigg(\sum_{n=2}^\infty
     \frac{\zeta(n)u^n}{n}\Big(2^n-1-(-1)^n\Big)\Bigg) \label{Fub}
\end{eqnarray}
where the first formula is applicable if $-1<u<1/2$, the second if $|u|<1/2$
and where $\gamma_E$ denotes the Euler-Mascheroni constant.

Adopting a strictly perturbative approach, we can use the original
result leading to (\ref{Vq}) (or in other words re-expand
$\alpha_{\overline{\Rsub MS}}({\bf q}^2)$) to find the potential in
position space. From
\begin{eqnarray}
  \alpha_{\Rsub V}({\bf q}^2) & = & \alpha_{\overline{\Rsub MS}}(\mu^2)\Bigg(
     1 + \frac{\alpha_{\overline{\Rsub MS}}(\mu^2)}{4\pi}\Big( -\beta_0\ln
      \frac{\bf q^2}{\mu^2} + a_1\Big) \nonumber \\
  && \hspace{-2em} +
     \Big(\frac{\alpha_{\overline{\Rsub MS}}(\mu^2)}{4\pi}\Big)^2 \Big(
     (\beta_0\ln\frac{\bf q^2}{\mu^2})^2 - (2\beta_0a_1+\beta_1)\ln
     \frac{\bf q^2}{\mu^2} + a_2\Big) + \ldots \Bigg),
\end{eqnarray}
where $\mu$ is the renormalization scale, we see that we need the Fourier
transform of $\ln^m(\mu^2/{\bf q^2})$ which we easily obtain from ${\cal F}$
because
$$
   \ln^m\frac{\mu^2}{\bf q^2} = \Bigg[\frac{\partial^m}{\partial u^m}
   \Big(\frac{\mu^2}{\bf q^2}\Big)^u\Bigg]\Bigg|_{u=0}
$$
and thus
\begin{equation}
    \int\!\!\frac{d^3q}{(2\pi)^3}\ln^m\frac{\mu^2}{\bf q^2}
    \frac{e^{i{\bf qr}}}{\bf q^2} = \Bigg(\frac{\partial^m}{\partial u^m}
    {\cal F}(r,\mu,u)\Bigg)\Bigg|_{u=0} \label{lnq}.
\end{equation}
Hence the potential in position space becomes
\begin{eqnarray}
 V(r) & = & -C_F\frac{\alpha_{\overline{\Rsub MS}}(\mu^2)}{r}\Bigg(
      1 + \frac{\alpha_{\overline{\Rsub MS}}(\mu^2)}{4\pi}\Big(
          2\beta_0\ln(\mu r^\prime)+a_1\Big) \nonumber \\
   && + \Big(\frac{\alpha_{\overline{\Rsub MS}}(\mu^2)}{4\pi}\Big)^2 \Big(
      \beta_0^2(4\ln^2(\mu r^\prime)+\frac{\pi^2}{3}) \label{Vr} \\
   && \quad +2(\beta_1+2\beta_0a_1)\ln(\mu r^\prime)+a_2\Big) + \ldots \Bigg)
      \nonumber
\end{eqnarray}
with $r^\prime\equiv r\exp(\gamma_E)$. One should note that going beyond the
strictly perturbative approach is quite difficult and may lead to
inconsistencies as the running coupling has a pole and thus the Fourier
transform of (\ref{Vq}) does not exist. But as the pole is an artifact which
arises because a perturbative expression is extrapolated deep into the
non-perturbative regime, it should not be taken too seriously.

\begin{figure}\begin{center}
 \epsfxsize 12cm \mbox{\epsfbox{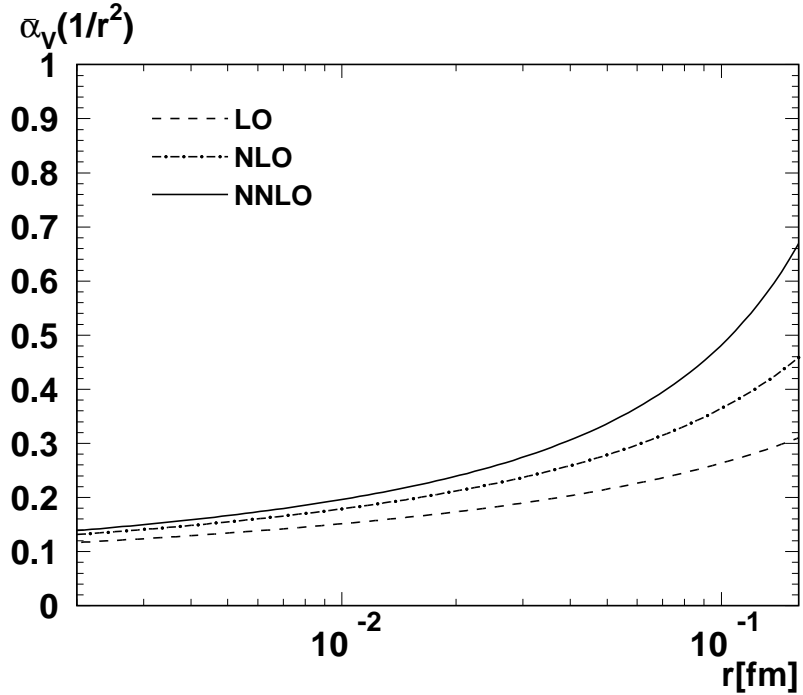}}
 \end{center}
 \caption{The potential (times $-r/C_F$) in position space at different orders
   choosing $\mu=1/r$. The range in $r$ roughly corresponds to the range in
   $|{\bf q}|$ displayed in the previous figures: $r=10^{-2.6}$fm $\approx
   0.002$fm $\sim 100$GeV and $r=10^{-0.8}$fm $\approx 0.16$fm $\sim
   1$GeV.\label{fig:rvr}}
\end{figure}

The result (\ref{Vr}) may be exploited in several ways: we certainly would not
use it as it stands because of the possibly large logarithms, but choose a
scale $\mu$ that reduces the higher order corrections. The first and in some
sense natural choice is $\mu_1=1/r$, leading to the curves displayed in Fig.\ 
\ref{fig:rvr} where the effect of the loop-corrections on $rV(r)$ is shown,
using the same input parameters as before. As the tree-level prediction would
correspond to a constant in this plot, it has been omitted.  The graph is
essentially the same as Figure \ref{fig:ava} and therefore will not be
discussed further.

A second choice motivated by noticing that due to (\ref{Fub}), $\mu$ will
always appear in combination with $r^\prime$, is $\mu_2=1/r^\prime$. This
way we remove all terms involving $\gamma_E$ form the coefficients. Note that
these terms are not small numerically because in $n$th order they involve
$(2\gamma_E\beta_0)^m\approx\beta_0^m$ for all $m\le n$ as well as
similar contributions from the other $\beta_m$.

A third choice, which is frequently encountered in other contexts as well,
would be to select $\mu$ in such a way as to remove the first-order coefficient
completely, i.e.\ $\mu_3(r)=\exp[-\gamma_E-a_1/(2\beta_0)]/r$, which leads to
\begin{equation}
  V_3(r) = -C_F\frac{\alpha_3}{r}\Bigg(1+\Big(\frac{\alpha_3}{4\pi}\Big)^2
   \Big(a_2-a_1^2-\frac{\beta_1}{\beta_0}a_1
   +\frac{(\pi\beta_0)^2}{3}\Big)\Bigg)
\end{equation}
with $\alpha_3\equiv\alpha_{\overline{\Rsub MS}}(\mu^2_3(r))$. This approach
is in effect similar to defining an effective charge like in QED, but one
should remember that it cannot be interpreted as a Dyson summation of the
one-loop vacuum polarization and therefore there is no reason why it should
constitute a superior choice. A real Dyson summation would lead to problems
with gauge invariance except for some part of the fermion loop contribution.

Yet another possibility, which is similar in spirit, would be to choose $\mu$
in such a way as to remove all $n_f$-dependence from the coefficients $a_n$,
i.e.\ to use the BLM scale setting prescription \cite{BLM}. This procedure
leads to
\begin{equation}
  V_{\rm _{BLM}}(r) = -\frac{C_F}{r}\Bigg(\alpha_{\overline{\Rsub MS}}\Big(
     \frac{1}{r_1^2}\Big)+a_1^{\mbox{\tiny BLM}}
     \frac{\alpha^2_{\overline{\Rsub MS}}(1/r_2^2)}{4\pi}
     +a_2^{\mbox{\tiny BLM}}
      \frac{\alpha^3_{\overline{\Rsub MS}}(1/r^{\prime2})}{(4\pi)^2}\Bigg)
\end{equation}
with
\begin{eqnarray}
  a_1^{\mbox{\tiny BLM}} & = & -\frac{8}{3}C_A \\
  a_2^{\mbox{\tiny BLM}} & = & \Big(\frac{133-396\zeta_3}{9}+\frac{24\pi^2
     -\pi^4}{4}\Big)C_A^2 - \frac{385-528\zeta_3}{12}C_AC_F \\
  r_1^2 & = & r^{\prime2} e^{5/3}\\
  r_2^2 & = & r^{\prime2} \exp\Big(
     \frac{434-504\zeta_3}{192}-\frac{315-432\zeta_3}{192}\frac{C_F}{C_A}
     \Big)\approx r^{\prime2} e^{-0.42}.
\end{eqnarray}

Finally, a fifth choice follows from proceeding in analogy to momentum space
and defining
\begin{equation}
  V(r) = -C_F\frac{\bar\alpha_{\Rsub V}(1/r)}{r}.
\end{equation}
From (\ref{Vr}) one may calculate the $\beta$-function of the new
coupling, with the result
\begin{equation}
  \bar\beta_2^{V} = \beta_2^V + \frac{\pi^2}{3}\beta_0^3,
\end{equation}
and, after determining the initial value for $\bar\alpha_{\Rsub V}$ at $M_Z$
form (\ref{Vr}) (where there are again different choices for $\mu$ possible,
we have used $\mu=1/r$) evolve it to the right distance. It is evident that
the appearance of $\gamma_E$ and the $\zeta$-functions makes $\bar\alpha_{
\Rsub V}$ differ from $\alpha_{\Rsub V}$, and consequently the same
statements holds for the $\beta$-functions.

\begin{figure}\begin{center}
 \epsfxsize 12cm \mbox{\epsfbox{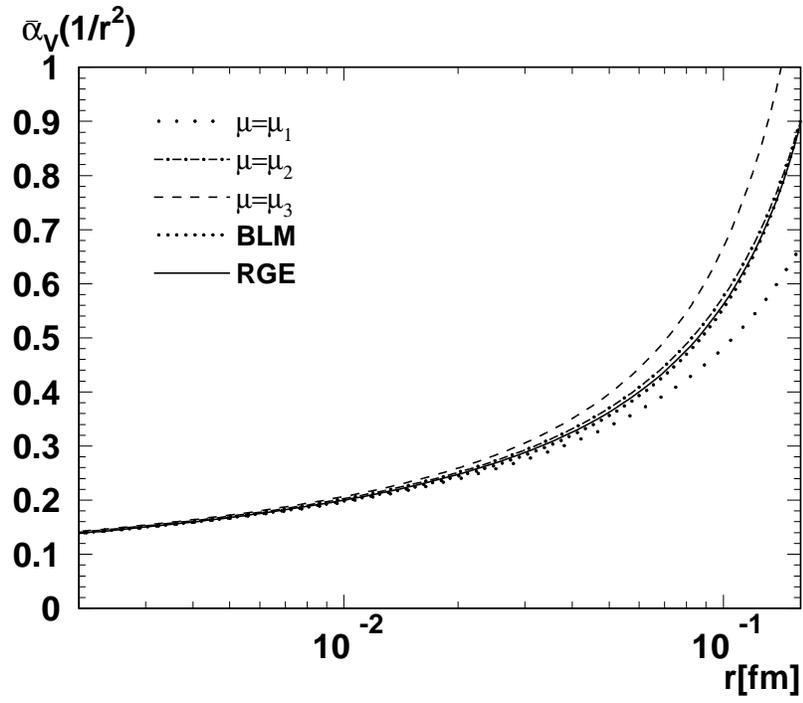}}
 \end{center}
 \caption{$-rV(r)/C_F$ in the different schemes described in the
  text.\label{fig:vschemes}}
\end{figure}

\begin{figure}\begin{center}
 \epsfxsize 12cm \mbox{\epsfbox{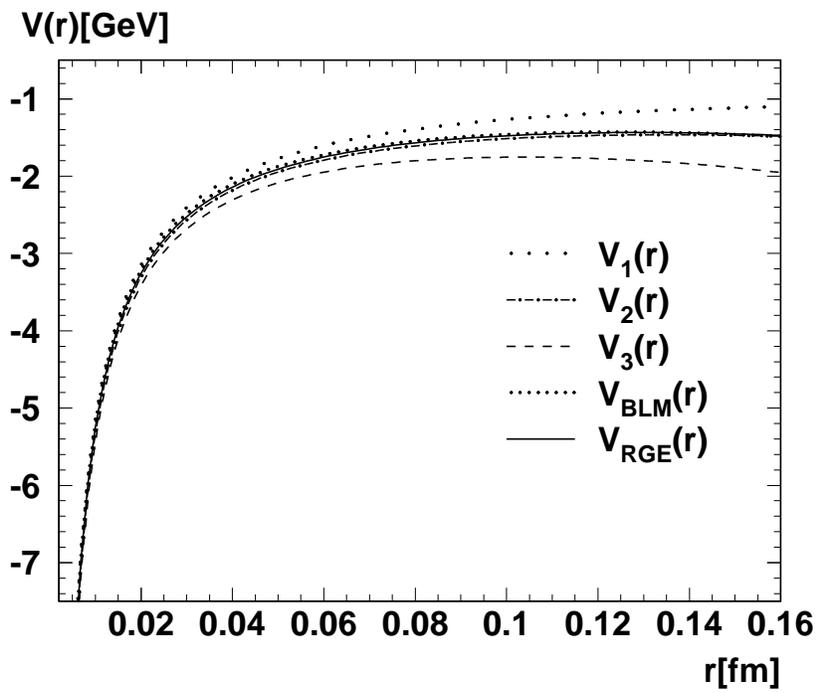}}
 \end{center}
 \caption{Three-loop potential in the different schemes described in the
  text.\label{fig:vschemesr}}
\end{figure}

The five approaches obviously result in different predictions for the
three-loop potential --- the difference formally being of the order $\alpha^4$
--- as is explicitly demonstrated in Figures \ref{fig:vschemes} and
\ref{fig:vschemesr}. One should note that there is no a priori reason to
prefer any choice of scale. The second choice $\mu=1/r^\prime$ might be an
exception, as it consistently absorbs terms which only arise from the Fourier
transformation and are not directly related to the dynamics. It suggests that
the distance complementary to $|\bf q|$ is not $r$ but $r^\prime$. But still
``non-dynamical'' terms $\zeta_n$ remain even in this scheme. The
BLM-prescription, which requires more than a single scale, is also physically
motivated, and the nice agreement between the two curves and the curve
resulting from the renormalization group evolution (labelled ``RGE'' in the
figures) may serve as an argument in favour of these approaches.

The coupling $\bar\alpha_{\Rsub V}$ is already quite large at a distance
$r\sim$ 0.5GeV$^{-1}$ or 0.1fm, and its size is even strongly increased in the
second and third approach due to a reduction of its scale implying that the
non-perturbative regime starts at smaller separations. For five flavours for
example we have $\mu_2\approx0.56/r$ and $\mu_3\approx0.41/r$. As a
consequence there is a large scheme-dependence of the potential in the region
$r>0.07$fm. From Figure \ref{fig:vschemesr} we see that the situation is even
worse as the perturbation series must break down near $r=0.1$fm because the
potential starts to bend down there already. We thus should not trust the
perturbative result for distances larger than about 0.08fm, which is
consistent with the assumption that the perturbative potential should be
reliable if $r\Lambda_{\Rsub QCD} < 0.07\ldots0.1$ is satisfied%
\footnote{For five flavours and $\alpha_{\Rsub \overline{MS}}(M_Z)=0.118$ we
  find $\Lambda_{\Rsub QCD}\approx210$MeV.} \cite{BT81,KMP86}.

An interesting question is whether our result gives any indication on the onset
of a linear rise of the potential, or in other words on the existence of
confinement. Unfortunately, the situation seems unclear in this respect.  As
Figure \ref{fig:rvr} proves, each new term of the perturbation series that is
added makes the deviation from a pure Coulomb prediction larger and the
potential more attractive. But according to the conventional definition
(\ref{QCDV}), a linearly {\em rising} potential requires a {\em negative}
coupling, the perturbative curve for $\bar\alpha_{\Rsub V}$ thus tends into the
wrong direction, and there is no sign of a nontrivial zero of this function.
The only indication on the existence of a zero is the breakdown of perturbation
theory at distances of the order $r\approx0.1$fm (or in other words the
presence of the Landau pole) and the fact that, with increasing number of
loops, this point appears at smaller and smaller distances. In some sense the
gap between the perturbative and the linear part thus becomes broader.

\section{Conclusion}

We have presented results of a full NNLO-calculation of the perturbative
static potential in QCD both in momentum and position space, including a
description of the technique employed, and we have argued that the respective
couplings $\alpha_{\Rsub V}$ and $\bar\alpha_{\Rsub V}$ are universal in the
sense that they describe the potential for other infinitely heavy colour
sources such as static gluinos as well. The $\beta$-functions corresponding to
the two couplings have been derived. The main result is that the two-loop
contribution is nearly as large as the one-loop term, both make the interaction
more attractive, and that the perturbative potential seems to be reasonably
reliable up to distances $r\Lambda_{\Rsub QCD}<0.07$. For larger
values a strong scale-dependence remains and the perturbation series even
breaks down above about 0.1fm.
Although the results may be interpreted as an indication of a general
tendency of higher loop corrections to strengthen the force between a
quark-antiquark pair, to take them as a proof of confinement would be going
too far.

Starting from the results presented in this paper, it would be important to
know, first, the impact of the two-loop contribution to the potential as a
whole and, second, of the ambiguity due to different choices of scale on the
energy levels and decay widths of quarkonia. The first point could be
investigated in the toponium system as this is only sensitive to very short
distances and does not suffer from the uncertainties arising from
intermediate and large separations. The second point, however, would require
the specification of some potential model for the latter region and thus the
analysis would depend on additional parameters. Nevertheless, a revised
study of the type performed in \cite{KMP86} would be very interesting.

\subsection*{Acknowledgments}
The author would like to thank Prof.\ J. H. K\"uhn for carefully reading the
manuscript and his continuous comments and advice, and Prof.\ Th. Mannel
for many interesting discussions.

\section*{Appendix: Calculation of the two-loop integrals}
When trying to compute the two-loop integrals $I(a,b,c;1,1)$, it is convenient
to combine gluon propagators using the standard formula for Feynman parameters
and to combine gluon and source propagators using the modified version
$$
  \frac{1}{a^m b^n} = \frac{\Gamma(n+m)}{\Gamma(n)\Gamma(m)}
    \int_0^\infty\!d\alpha\frac{\alpha^{m-1}}{(a\alpha+b)^{n+m}}.
$$
With this technique the two momentum integrations can be performed one after
the other, and after some rescalings of the $\alpha$-parameters one is left
with
\begin{equation} \label{eq:I}
  I(a,b,c;1,1) = 
    \Big(4\pi\mu^2)^{2\epsilon}(-q^2)^{D-\Sigma}\frac{2\Gamma(\Sigma-D)}{
    \Gamma(a)\Gamma(b)\Gamma(c)}\tilde I(a,b,c)
\end{equation}
where $\Sigma=a+b+c+1$ and
\begin{eqnarray*}
 \tilde I(a,b,c) &=&
   2\frac{\Gamma(\Sigma+1-D)}{\Gamma(\Sigma-D)}\int_0^1\!\!dx\;dy
    \int_0^\infty\!\!d\alpha\;d\beta\;
    x^{\frac{D}{2}-b-2}(1-x)^{\frac{D}{2}-c-1} \\ &&
    \times y^{b+c-\frac{D}{2}-1}(1-y)^{a-1}
     \Bigg[(\alpha+\beta)^2+ \alpha^2\frac{1-x}{xy}+y(1-y)\Bigg]^{D+1-\Sigma}.
\end{eqnarray*}
Making the change of variables $\alpha = u\rho$ and $\beta=(1-u)\rho$, the
two integrations corresponding to $\alpha$ and $\beta$ can be done, with the
result
\begin{equation} \label{eq:It}
   \tilde I =
   \int_0^1\!\!dx\;dy\;x^{\frac{D-3}{2}-b}(1-x)^{\frac{D-3}{2}-c}
   y^{\frac{D-3}{2}-a}(1-y)^{D-2-b-c}\arctan\sqrt{\frac{1-x}{xy}}.
\end{equation}
The presence of the $\arctan$-function makes it impossible (at least for the
author) to compute the remaining integral exactly for all values of $a,b,c$
and $\epsilon=(4-D)/2$. Hence only the cases really needed and only the
expansion in $\epsilon$ to the order required were treated.

For $I(1,1,1;1,1)$ this task is quite straightforward as $\tilde I(1,1,1)$ is
in fact finite in the limit $\epsilon\to0$ and its expansion in $\epsilon$ can
be computed without encountering any problems. But it should be noted that due
to (\ref{eq:I}) $\tilde I$ must be multiplied by $1/\epsilon$ to obtain $I$ and
thus the latter integral is divergent. It should also be mentioned that the
result quoted in section \ref{sec:Tech} has not been derived in the way just
described, but by constructing an equation involving the integral. This
method will be explained at the end of the appendix.

The problem with the other two integrals $\tilde I$ is that they do not exist
if we take $\epsilon=0$ and that it is difficult to factor out the divergence.
As a first step in this direction the substitution
$$ x = \frac{1}{1+t^2y} $$
can be applied to turn (\ref{eq:It}) into
\begin{equation}\label{Ii}
   \tilde I(i,1,2) = \int_0^\infty\!dt\;t^{-2-2\epsilon}\arctan t\cdot K_i(t^2)
\end{equation}
with
\begin{eqnarray*}
  K_1(t^2) & = & \int_0^1\!dy\;y^{-1-2\epsilon}(1-y)^{-1-2\epsilon}
    (1+t^2y)^{2\epsilon} \\
  & = & \frac{\Gamma^2(-2\epsilon)}{\Gamma(-4\epsilon)}
     F(-2\epsilon,-2\epsilon,-4\epsilon;-t^2) \\
  K_2(t^2) & = & \int_0^1\!dy\;y^{-2-2\epsilon}(1-y)^{-1-2\epsilon}
    (1+t^2y)^{2\epsilon} \\
  & = & \frac{\Gamma(-2\epsilon)\Gamma(-1-2\epsilon)}{\Gamma(-4\epsilon)}
     F(-2\epsilon,-1-2\epsilon,-1-4\epsilon;-t^2)
\end{eqnarray*}
and where $F(a,b,c;x)$ denotes the hypergeometric function. Although this
result may look quite impractical, it in fact helps: one factor $1/\epsilon$
has been factored out, and the expansion of the two hypergeomtric functions is
also possible up to terms which are not needed anyhow. Let us first examine
$K_1$:
\begin{eqnarray*}
  \lefteqn{F(-2\epsilon,-2\epsilon,-4\epsilon;x)} && \\
  & = & 1-\epsilon\sum_{n=1}^\infty
     \Bigg(\prod_{j=1}^{n-1}\frac{(j-2\epsilon)^2}{j-4\epsilon}\Bigg)
     \frac{x^n}{n!}
    = 1-\epsilon\sum_{n=1}^\infty\Bigg(\prod_{j=1}^{n-1}\Big(j+{\cal O}
     (\epsilon^2)\Big)\Bigg) \frac{x^n}{n!} \\
  & = & 1-\epsilon\sum_{n=1}^\infty\frac{x^n}{n} + {\cal O}(\epsilon^3)
    = 1+\epsilon\ln(1-x) + {\cal O}(\epsilon^3).
\end{eqnarray*}
Thus we find
$$ K_1(t^2) = 1+\epsilon\ln(1+t^2)+{\cal O}(\epsilon^3)k_1(t^2) $$ where the
residual function $k_1(t^2)$ is well behaved for $t\to0$. A similar
trick can be applied to $K_2$ if we first employ the relation \cite{Grad}
$$ F(a,b,c;x) = (1-x)^{-b}F\Big(b,c-a,c;\frac{x}{x-1}\Big)$$
which transforms $K_2$ into
$$
  K_2(t^2) = \frac{\Gamma(-2\epsilon)\Gamma(-1-2\epsilon)}{\Gamma(-4\epsilon)}
    (1+t^2)^{1+2\epsilon}F\Big(-1-2\epsilon,-1-2\epsilon,-1-4\epsilon;
    \frac{t^2}{1+t^2}\Big).
$$
Now
\begin{eqnarray*}
 \lefteqn{F(-1-2\epsilon,-1-2\epsilon,-1-4\epsilon;x)} && \\
 & = & 1-\frac{(1+2\epsilon)^2}{1+4\epsilon}x
       +\epsilon\frac{(1+2\epsilon)^2}{1+4\epsilon}\sum_{n=1}^\infty
     \Bigg(\prod_{j=1}^{n-1}\frac{(j-2\epsilon)^2}{j-4\epsilon}\Bigg)
     \frac{x^{n+1}}{(n+1)!} \\
 & = & 1-\frac{(1+2\epsilon)^2}{1+4\epsilon}x
       +\epsilon\frac{(1+2\epsilon)^2}{1+4\epsilon}\sum_{n=1}^\infty
       \frac{x^{n+1}}{n(n+1)} + {\cal O}(\epsilon^3) \\
 & = & 1-x+\epsilon(1-4\epsilon)x +\epsilon(1-x)\ln(1-x)+ {\cal O}(\epsilon^3).
\end{eqnarray*}
By noting that $F(0)=1$ and $F(a,a,c;1)=\Gamma(c)\Gamma(c-2a)/\Gamma^2(c-a)$
we can even improve the above expansion as $F$ must be of the form
$F=1-x+\Gamma(-1-4\epsilon)/\Gamma^2(-2\epsilon)x+\phi(x)$ where $\phi(x)$
vanishes for both $x=0$ and $x=1$. Thus
$$
  K_2(t^2) = (1+t^2)^{2\epsilon}\Bigg[
    \frac{\Gamma(-2\epsilon)\Gamma(-1-2\epsilon)}{\Gamma(-1-4\epsilon)}
    \Big(1-\epsilon\ln(1+t^2)\Big)-\frac{t^2}{1+2\epsilon} \Bigg]
    + {\cal O}(\epsilon^3)k_2(t^2)
$$
where $k_2$ vanishes at the origin and at infinity.

With these approximate formulae for the $K_i$ we can calculate the two missing
integrals from (\ref{Ii}) up to terms of the order ${\cal O}(\epsilon^2)$. The
divergences still present can be extracted by writing
\begin{equation}\label{DT}
  \int_0^\infty\!dt\;t^{-1-2\epsilon} f(t) = -\frac{f(0)}{2\epsilon}
  + \int_0^\infty\!dt\;t^{-1-2\epsilon}\Big( f(t)-f(0)\Theta(1-t)\Big)
\end{equation}
where the function $f(t)$ should be regular at the origin. The new integral is
finite in the limit $\epsilon\to0$, and consequently may be expanded. One
should note that in the case of $I(2,1,2)$ there is a term
$t^2(1+t^2)^{2\epsilon}$ in $K_2$ that removes the bad behaviour of the
integrand at the origin, but introduces the same kind of divergence for large
$t$. The trivial change of variables $t\to1/t$ reduces this term to the above
case again. It is essential that we treat the whole combination as of
order $1$: suppose, for example, we would expand $t^2(1+t^2)^{2\epsilon}$,
generating $2\epsilon t^2\ln(1+t^2)$. Inserted into (\ref{Ii}) and proceeding
as described this would produce an integral of the form
$$
  2\epsilon\int_0^\infty\!dt\; t^{-1+2\epsilon}\ln t\frac{\arctan t}{t}
$$
which is not of ${\cal O}(1)$ as we would assume, but of ${\cal
  O}(1/\epsilon)$. The term also demonstrates that it is crucial to know
that the omitted functions $k_i$ grow at most logarithmically after expansion
in $\epsilon$ and thus do not invalidate our power-counting in this
parameter.

Let us now come back to $I(1,1,1;1,1)$. Starting from the original definition
given in section \ref{sec:Tech}, shifting first $p\to p+r+q$, using $vq=0$
and
$$
   \frac{1}{(pv+rv+i\varepsilon)(rv+i\varepsilon)}=
   \frac{1}{pv+i\varepsilon}\Bigg[\frac{1}{rv+i\varepsilon}
      -\frac{1}{pv+rv+i\varepsilon}\Bigg]
$$
and then replacing $r\to-r$ in the first and shifting $r\to r-p-q$ in the
second term, one obtains:
\begin{eqnarray*}
  \lefteqn{ I(1,1,1;1,1) } && \\ &=&
  \int\!\widetilde{dp}\widetilde{dr}\frac{1}{(-p^2)(-r^2)(-(r+p+q)^2)}
  \frac{1}{(pv+rv+i\varepsilon)(rv+i\varepsilon)} \\ &=&
  -\int\!\widetilde{dp}\widetilde{dr}\frac{1}{(-p^2)(-r^2)(-(r-p-q)^2)}
  \frac{1}{pv+i\varepsilon} 
   \Bigg[\frac{1}{rv-i\varepsilon}+\frac{1}{rv+i\varepsilon}\Bigg].
\end{eqnarray*}
Hence
\begin{eqnarray*}
  \lefteqn{ I(1,1,1;1,1) } && \\ &=&
  \frac{1}{3}\int\!\widetilde{dp}\widetilde{dr}\frac{1}{
    (-p^2)(-r^2)(-(r-p-q)^2)(pv+i\varepsilon)}\Big[\frac{1}{rv+i\varepsilon}
    -\frac{1}{rv-i\varepsilon}\Big] \\ 
  &=& -\frac{2\pi}{3}i\int\!\widetilde{dp}\widetilde{dr}\frac{\delta(rv)}{
    (-p^2)(-r^2)(-(r-p-q)^2)(pv+i\varepsilon)} \\ 
  & = & \frac{2\pi}{3}i\sqrt\pi G(1,1;-1,-\frac{1}{2})\int\!\widetilde{dr}
    \frac{\delta(rv)}{(-r^2)(-(r-q)^2)^{1/2+\epsilon}}.
\end{eqnarray*}
In the last expression we can replace $i\pi\delta(rv)$ by
$-1/(rv+i\varepsilon)$ (due to $vq=0$ the principal value part of the
resulting integral vanishes) and use (\ref{three:one}) again, or we can
calculate the effective $D-1$-dimensional integral directly. Via both routes
we find the result given in section \ref{sec:Tech}.

\end{document}